\documentclass{article}
\usepackage{amsmath}
\usepackage{amsthm}

\usepackage{verbatim}
\usepackage{amssymb}
\usepackage{algorithm}
\usepackage{algorithmic}

\usepackage{caption}
\usepackage{graphicx} 

\usepackage[preprint]{neurips_2026}

\usepackage[utf8]{inputenc} % allow utf-8 input
\usepackage[T1]{fontenc}    % use 8-bit T1 fonts
\usepackage{hyperref}       % hyperlinks
\usepackage{url}            % simple URL typesetting
\usepackage{booktabs}       % professional-quality tables
\usepackage{amsfonts}       % blackboard math symbols
\usepackage{nicefrac}       % compact symbols for 1/2, etc.
\usepackage{microtype}      % microtypography
\usepackage{xcolor}         % colors

\title{Multi-Agent Reinforcement Learning in Markets with Congestion}
\workshoptitle{Economics for Machine Learning}

\author{%
  Qixuan Zai, Randall Berry\\
  Department of Electrical and Computer Engineering\\
  Northwestern University, Evanston, IL 60208, USA\\
  \texttt{Emails:\{QixuanZai2028@u.,rberry@\}northwestern.edu} \\
}

\begin{document}

\maketitle

\begin{abstract}
This paper investigates multi-agent reinforcement learning (MARL) in settings where firms compete for customers using congestible resources. We consider Bertrand competition in which firms compete by announcing prices and customers choose among firms based on both price and congestion.  The relationship between price, congestion and the quantity of customers willing to accept service is governed by an unknown inverse demand curve, which firms must learn through experience. Each firm is modeled as a self-interested learning agent that chooses its price to maximize profit. A growing literature has shown that independently learning MARL agents can develop tacitly collusive behavior. We examine how such behavior emerges in markets with congestible resources. Our results provide insight into how learning dynamics, state representation, and strategic interaction jointly shape competition, with implications for both economic learning and the design of learning-enabled markets.
\end{abstract}

\section{Introduction}
Many markets involve firms that compete for customers while sharing resources whose quality depends on the level of usage. Examples include cloud computing, where competing providers allocate capacity subject to congestion and performance constraints. In such markets, a firm's competitive decisions affect not only the prices customers face but also the quality of service available through the shared resource. These congestion externalities create strategic interactions that are absent from conventional models of price competition.

Existing congestion-based market models generally assume that market demand is known to all competing firms, enabling analytical characterization of equilibrium strategies~\cite{johari2004efficiency}. In practice, however, firms rarely possess complete knowledge of demand, which evolves over time as user preferences, competing technologies, and market conditions change~\cite{denboer2015dynamic}. We instead formulate repeated Bertrand competition over shared resources as a MARL problem in which competing firms adapt their pricing strategies in the face of unknown demand. While a linear inverse demand function is assumed to derive analytical benchmarks and characterize equilibrium behavior, the learning agents are not given this demand model. Instead, using deep Q-learning, they learn pricing strategies solely through repeated interactions with the environment. We investigate whether competing firms converge to competitive or collusive outcomes and how these outcomes depend on the information available to learning agents and their exploration behavior.

\section{System Model}

We consider a shared-resource market consisting of $N$ firms. Our model extends the congestion-market framework of \cite{Q3,Q1,Q2} by replacing the assumption of complete demand information with repeated learning. While the market follows the same congestion and Wardrop equilibrium structure as prior work, competing firms do not know the underlying demand function and must instead learn effective pricing strategies from observed market outcomes.

 Firms, indexed by $i= 1,\ldots,N$, share access to a common resource with capacity $W_e$ and compete for a common pool of non-atomic customers. Let $x_{i}$ denote the amount of users served by firm $i$ and let $\mathbf{x} = (x_i)_{i=1}^N$. The quality of service experienced by users is determined by congestion on the resource, modeled through a convex increasing congestion cost 
 \begin{equation}   
    \hat{g}\left(\mathbf{x}\right) \triangleq g\left(\frac{X}{W_e}\right),
    \label{eq:congest_en}
\end{equation}
where $X = \sum_{i=1}^N x_i$.

Each firm collects a fixed service price $p_i$ from each user it serves, yielding a revenue of $x_i p_i$.\footnote{We consider models with any cost normalized to zero so that revenue is equivalent to profit.} The delivered price of a firm $i$ is the sum of its service price and the congestion of its service, i.e., $p_i + \hat g(\mathbf x)$ for $1\le i\le N$. 
Customer demand is characterized by an unknown inverse demand function $P(X)$, where $P(X)$ denotes the maximum delivered price that can be sustained when the total demand is $X$.
 Note that the demand function depends on the delivered price, which models customers that care about both the price they pay for service and the quality of service. The amount of users served by each firm and the announced prices satisfy the Wardrop equilibrium conditions, that is, the delivered prices of all firms that are serving users are the same and are the maximum price that can be accepted by all the customers serviced \cite{wardrop1952some}. If a firm $i$ is not serving any customers, then its delivered price must be greater than or equal to $P(X)$. 
 This can be expressed as 
\begin{equation}
\begin{aligned}
P(X) &\leq  p_i + \hat{g}(\mathbf{x}), \quad 1 \le i \le N,\\
P(X) &= p_i + \hat{g}(\mathbf{x}), \quad \text{for all $i$ with $x_i > 0$}
\end{aligned}
\label{eq:wardropeq}
\end{equation}

The market is characterized by an unknown inverse demand function $P(X)$, which is not revealed to the learning agents. In each interaction, providers observe the resulting market outcome but do not observe the underlying demand function or competitors' private decision process.

\subsection{Bertrand Congestion Competition}

Under Bertrand competition, service providers compete by selecting service prices. The resulting customer allocation is determined endogenously through the Wardrop equilibrium conditions in (\ref{eq:wardropeq}), which couple the announced prices, congestion costs, and total market demand. This model captures markets in which providers compete directly on price while customers choose the provider offering the lowest delivered price.

Classical Bertrand theory predicts that firms offering homogeneous products compete prices down to marginal cost, yielding the well-known Bertrand paradox~\cite{bertrand1883theorie,tirole1988theory}. In contrast, recent work has demonstrated that reinforcement-learning agents may instead converge to tacitly collusive pricing strategies without explicit communication~\cite{calvano2020algorithmic,klein2021autonomous,denicolo2023artificial}. A primary objective of this paper is to determine how parameters of learning algorithms in congestion-based markets influence the learned pricing behavior.

Each firm seeks to select its announced service price to maximize its revenue given by $x_ip_i$,
where the demand allocated to each firm depends on the prices announced by all competing providers together with the resulting congestion costs.\footnote{When multiple firms have the same delivered price, we assume that the demand is split evenly among them.}

\section{Multi-Agent Reinforcement Learning}

Now we consider a setting where each firm is a learning agent that repeatedly plays a Bertrand game with congestion as in the previous section.  
 At each round, all firms simultaneously select prices, which lead to a market equilibrium of users.  The immediate reward a firm obtains is then its revenue $r_i = p_ix_i$. The agents do not know the underlying demand function and must instead learn effective strategies solely through repeated interaction with the environment. Each agent's state consists of the joint actions (prices) taken during the previous $K$ rounds, providing a finite history of market behavior. The action space consists of a discrete set of allowable prices. Unlike much of the prior literature on algorithmic collusion, which typically uses tabular Q-learning~\cite{calvano2020algorithmic,klein2021autonomous}, we employ independent Deep Q-Network (DQN) agents~\cite{mnih2015human}, which extend more readily to larger or continuous state and action spaces.  Each agent maintains its own value network and replay buffer, while treating the remaining agents as part of the environment. Standard experience replay and target networks are used to stabilize training.

\section{Empirical Results}
In this section we give several numerical examples to illustrate the impact of the number of firms, the state representation and the exploration decay rate, $\delta$.  For all plots we normalize the capacity as $W_e =1$ and assume the inverse demand is given by $P(X) = 1-X$, with a discrete price set given by $0.05, 0.15, 0.25, \ldots, 0.95$.
Under these assumption, we derive the full-information Nash equilibrium and a collusive benchmark, where all firms coordinate to maximize total revenue.

\begin{figure*}[htbp]
\centering
\includegraphics[width=\linewidth]{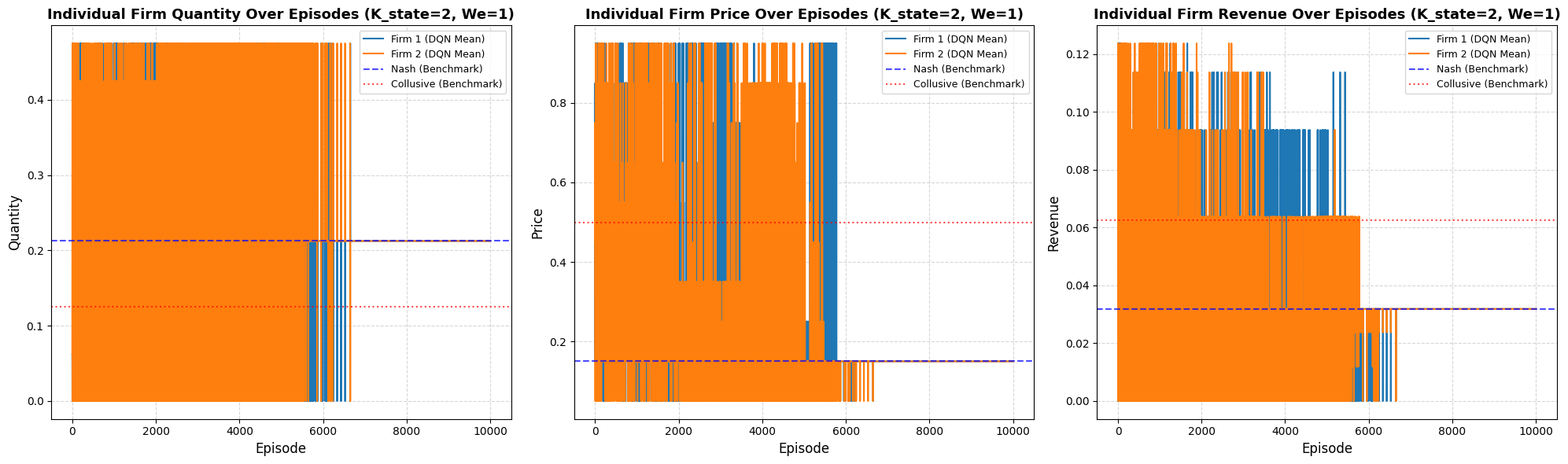}
\caption{Learning Dynamics for Bertrand Congestion Model over episodes when $W_e=1$, $N=2$, $K=2$, $\delta=0.99$.}
\label{fig:episodes}
\end{figure*}

With discrete prices, the Bertrand game will always have a Nash equilibrium where all firms select the lowest price ($0.05$), but may also have others.  In particular, for $N=2$ firms, there is also a Nash equilibrium where all firms select $p = 0.15$.  Figure~\ref{fig:episodes} shows an example of the learning dynamics for $N=2$ firms when the $K=2$ and $\delta = 0.99$.  In this example it can be seen that the firms eventually learn the Pareto-dominant Nash equilibrium with $p = 0.15$.

\begin{figure*}[htbp]
\centering
\includegraphics[width=\linewidth]{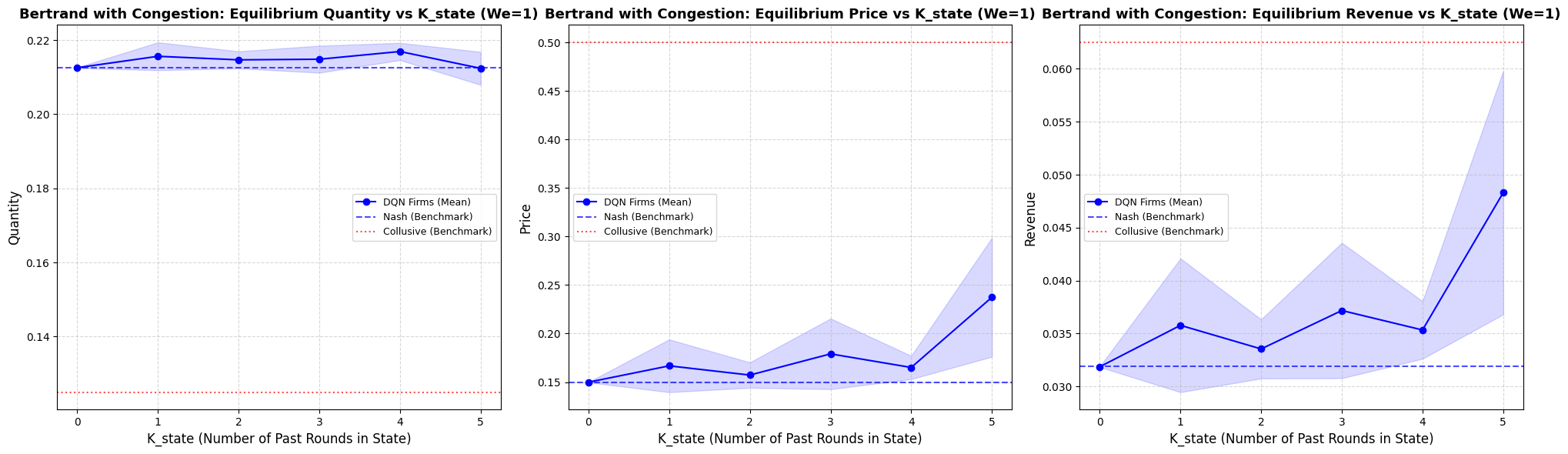}
\caption{Learning outcomes versus number of states for Bertrand Congestion Model with $W_e=1$, $N=2$, $\delta=0.99$.}
\label{fig:incumbent}
\end{figure*}

Figure~\ref{fig:incumbent} presents results to show the impact of changing the state representation $K$, where all other parameters are the same as in Figure~\ref{fig:episodes}.  Here, $K=0$ correspond to a setting where agents use no state information and so are essentially in a bandit setting.  In that case they always appear to find a Pareto-dominant Nash Equilibrium. As the state representation increases, they do sometimes learn to coordinate on high prices, but are well below the collusive benchmark.  The amount of collusion appears to increase as $K$ grows.

We note that convergence to the Pareto-dominant Nash equilibrium, even in the stateless (bandit) setting, already reflects a mild form of coordination: with multiple static Nash equilibria available, agents consistently select the one more favorable to firms rather than the fully competitive price $p=0.05$. This is distinct from the tacit collusion documented in the literature, which sustains prices above any static Nash equilibrium through implicit reward-punishment dynamics across repeated play. Our results with larger $K$ show collusion in this stronger sense, with prices exceeding the Pareto-dominant Nash benchmark.

\begin{figure*}[htbp]
\centering
\includegraphics[width=\linewidth]{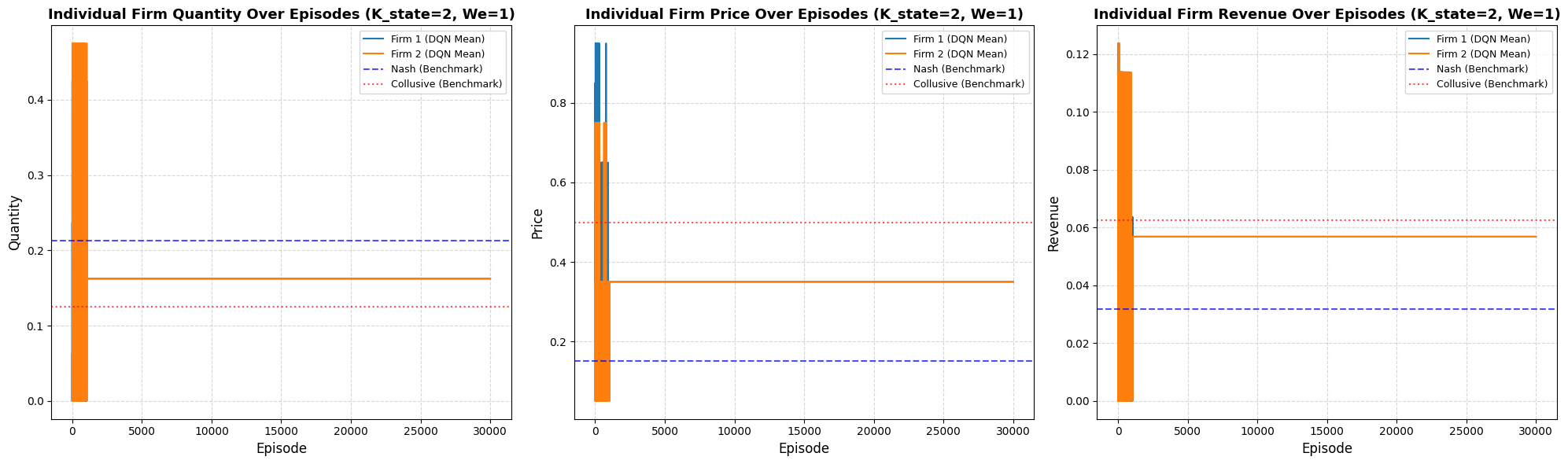}
\caption{Impact of Exploration Decay Rate on the Learning Dynamics for Bertrand Congestion Model over episodes when $W_e=1$, $N=2$, $K=2$, $\delta=0.9$.} 
\label{fig:algorithmt6}
\end{figure*}

Next, in Figure~\ref{fig:algorithmt6}, we show one result for a lower exploration rate ($\delta = 0.9$), where all other parameters are the same as in Figure~\ref{fig:episodes}. In this case the learned prices for both agents are around 0.35 which is greater than the Nash equilibrium benchmark, suggesting that they are learning to collude.

\begin{figure*}[htbp]
\centering
\includegraphics[width=\linewidth]{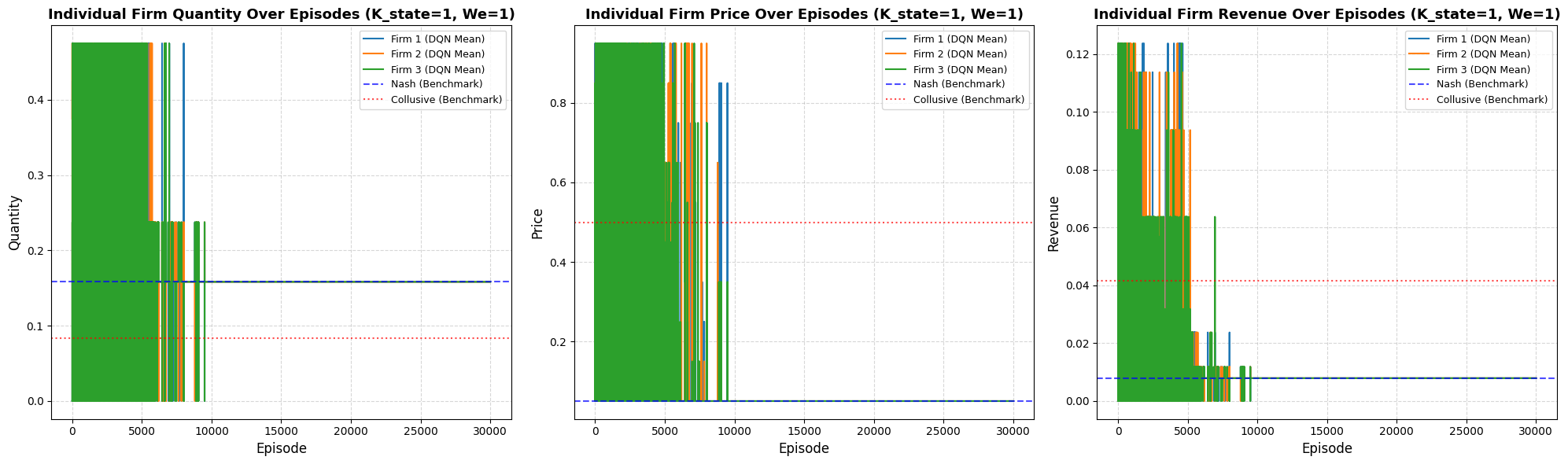}
\caption{Impact of The Number of Agents on the Learning Dynamics for Bertrand Congestion Model over episodes when $W_e=1$, $N=3$, $K=1$, $\delta=0.99$.} 
\label{fig:algorithm3}
\end{figure*}

Finally in Figure~\ref{fig:algorithm3} we show an example with $N=3$ agents, again with all other parameters the same as in Fig.~\ref{fig:episodes}. For the parameters we are using, with $N= 3$ agents, the only Nash equilibrium is for all agents to choose the smallest price ($p=0.05$), which is the outcome of this experiment. This is consistent with our analytical result that the Pareto-dominant Nash equilibrium from the 
$N = 2$ case does not survive at $N=3$, leaving only the fully competitive price, which is exactly what the learning agents converge to.\footnote{Lowering $\delta$, with $N=3$ can again lead to higher prices, we do not show this due to space limitations.}

\section{Conclusion}
This paper investigated how firms learn in congestion-based markets when market demand is unknown. We introduced a MARL framework for this setting and evaluated how adaptive learning algorithms navigate congestion externalities in Bertrand competition. We find evidence that the state representation, the exploration decay rate, and the number of firms all have an impact on the emergence of collusive prices, suggesting that market designers and regulators should weigh these factors when assessing collusion risk in learning-enabled markets. Future work could extend this analysis to alternative demand specifications and larger markets.

\clearpage
\bibliographystyle{plain}
\bibliography{ref}

@inproceedings{Q2,
  author    = {Y. Liu and Q. Zai and R. A. Berry},
  title     = {Wireless Market Competition with Time/Frequency Prioritized Spectrum Sharing},
  booktitle = {23rd International Symposium on Modeling and Optimization in Mobile, Ad Hoc, and Wireless Networks (WiOpt)},
  pages     = {1--8},
  year      = {2025}
}

@inproceedings{Q1,
  author    = {Q. Zai and Y. Liu and R. A. Berry},
  title     = {Poster: Comparison of Market Models for Time/Frequency Prioritized Spectrum Sharing},
  booktitle = {Proceedings of the 26th International Symposium on Theory, Algorithmic Foundations, and Protocol Design for Mobile Networks and Mobile Computing},
  pages     = {409--410},
  year      = {2025}
}

@inproceedings{Q3,
  author = {Q. Zai and Y. Liu and R. A. Berry},
  title  = {Comparison of Market Models for Time/Frequency Prioritized Spectrum Sharing},
  booktitle = {Proceedings of {EAI GameNets}},
  year   = {2026}

}

@article{klein2021autonomous,
  author  = {T. Klein},
  title   = {Autonomous Algorithmic Collusion: Q-learning Under Sequential Pricing},
  journal = {RAND Journal of Economics},
  volume  = {52},
  number  = {3},
  pages   = {538--558},
  year    = {2021}
}

@article{denicolo2023artificial,
  author  = {V. Denicol{\`o} and E. Calvano},
  title   = {Artificial Intelligence, Algorithmic Pricing and Collusion},
  journal = {European Economic Review},
  volume  = {153},
  pages   = {104386},
  year    = {2023}
}

@article{johari2004efficiency,
  author  = {R. Johari and J. N. Tsitsiklis},
  title   = {Efficiency Loss in a Network Resource Allocation Game},
  journal = {Operations Research},
  volume  = {52},
  number  = {3},
  pages   = {474--485},
  year    = {2004}
}

@article{bertrand1883theorie,
  author  = {J. Bertrand},
  title   = {Th{\'e}orie Math{\'e}matique de la Richesse Sociale},
  journal = {Journal des Savants},
  volume  = {67},
  pages   = {499--508},
  year    = {1883}
}

@book{tirole1988theory,
  author    = {J. Tirole},
  title     = {The Theory of Industrial Organization},
  publisher = {MIT Press},
  year      = {1988}
}

@article{calvano2020algorithmic,
  author  = {E. Calvano and V. Denicol{\`o} and M. Vincenzo and S. Pastorello},
  title   = {Artificial Intelligence, Algorithmic Pricing, and Collusion},
  journal = {American Economic Review},
  volume  = {110},
  number  = {10},
  pages   = {3267--3300},
  year    = {2020}
}

@article{wardrop1952some,
  author  = {J. G. Wardrop},
  title   = {Some Theoretical Aspects of Road Traffic Research},
  journal = {Proceedings of the Institution of Civil Engineers},
  volume  = {1},
  number  = {3},
  pages   = {325--362},
  year    = {1952}
}

@article{mnih2015human,
  author  = {V. Mnih et al.},
  title   = {Human-Level Control Through Deep Reinforcement Learning},
  journal = {Nature},
  volume  = {518},
  number  = {7540},
  pages   = {529--533},
  year    = {2015}
}

@article{denboer2015dynamic,
  author  = {A. V. den Boer},
  title   = {Dynamic Pricing and Learning: Historical Origins, Current Research, and New Directions},
  journal = {Surveys in Operations Research and Management Science},
  volume  = {20},
  number  = {1},
  pages   = {1--18},
  year    = {2015}
}
%%%%%%%%%%%%%%%%%%%%%%%%%%%%%%%%%%%%%%%%%%%%%%%%%%%%%%%%%%%%
\clearpage
\appendix

\end{document}